\documentclass{an}
\usepackage{graphicx}
\usepackage{times}
\usepackage{fancyhdr}
\usepackage{color}
\def\del#1{}
\def\aj{AJ}
\def\aap{A\&A}
\def\prd{PRD}

\newcommand{\ltsima}{$\; \buildrel < \over \sim \;$}
\newcommand{\lsim}{\lower.5ex\hbox{\ltsima}}
\newcommand{\gtsima}{$\; \buildrel > \over \sim \;$}
\newcommand{\gsim}{\lower.5ex\hbox{\gtsima}}

\newcommand{\eps}{\varepsilon}

\def\RM{{\rm RM}}
\def\lesssim{\mathrel{\hbox{\rlap{\hbox{\lower4pt\hbox{$\sim$}}}\hbox{$<$}}}}
\def\gtrsim{\mathrel{\hbox{\rlap{\hbox{\lower4pt\hbox{$\sim$}}}\hbox{$>$}}}}

\def\vx{\vec{x}}
\def\vr{\vec{r}}
\def\vxp{\vec{x}_\perp}
\def\vrp{\vec{r}_\perp}
\def\vk{\vec{k}}
\def\vkp{\vec{k}_\perp}

\def\RM{{\rm RM}}
\def\obs{{\rm obs}}

\def\L0{\Lambda_{0}}

\begin{document}

\title{Future magnetic fields studies using the Planck surveyor experiment}

\author{Torsten A. En{\ss}lin\inst{1} \and Andr{\'e}
  Waelkens\inst{1} \and Corina Vogt\inst{1,2} \and Alexander A. Schekochihin\inst{3}} 
\institute{Max-Planck-Institute f\"ur Astrophysik, Karl-Schwarzschild-Str. 1,
  85741 Garching, Germany 
\and
Stichting ASTRON, P.O. Box 2, NL--7990 AA Dwingeloo, The Netherlands
\and
DAMTP/CMS, University of Cambridge,
Wilberforce Road,
Cambridge CB3 0WA, United Kingdom
}

\date{Received; accepted; published online}

\abstract{The Planck mission will permit measurements of the polarization of
  the cosmic microwave background and of polarized foregrounds such as our own
  Galaxy with an unprecedented combination of accuracy and completeness. This
  will provide information on cosmological and galactic magnetic fields. The
  latter can be studied in detail via nearly Faraday-rotation free synchrotron
  and polarized dust emission. Methods are discussed to extract physically relevant
  information on the magnetic turbulence from Planck data and other
  measurements.  }

\correspondence{ensslin@mpa-garching.mpg.de}

\maketitle

\section{The Planck surveyor mission}
\subsection{The experiment}
``Planck is a mission of the European Space Agency designed to answer key
cosmological questions.  Its ultimate goal is to determine the geometry and
content of the Universe, and which theories describing the birth and evolution
of the Universe are correct.  To achieve this ambitious objective, it will
observe the Cosmic Microwave Background radiation (CMB), emitted about 13
thousand million years ago, just 400,000 years after the Big Bang.  Today the
CMB permeates the Universe and is observed to have an average temperature of
2.726 K.  Small deviations from this average value (the so-called
anisotropies), observable at angular scales larger than a few arcminutes, encode
a wealth of information on the properties of the Universe in its infancy.  The
objective of Planck is to measure these properties with an unprecedented
accuracy and level of detail.'' (\cite{PlanckBluebook2005}).

The Planck satellite will operate at the 2$^{\rm nd}$ Lagrange point of the
Sun-Earth system, with both Sun and Earth in the direction of the axis of the
satellite, around which it rotates with 1 rotation per minute. Two sets of
detectors will be on board, the HEMT receivers for the Low Frequency Instrument
(LFI) ranging from 30 to 70 GHz and bolometers for the High Frequency
Instrument (HFI) ranging from 100 to 860 GHz. Both instruments will have
receivers sensitive to linear polarization.

The instrument beams will have resolutions of 5 to 30 $\rm arcmin$ and observe
the sky with an inclination of $85^{\circ}$ with respect to the spin
axis. About every 60 rotations, the spin axis will be rearranged to a new direction
within a $10^\circ$ cone pointing towards the Sun. Within seven months,
the complicated scanning strategy of Planck will have covered the full sky at
least once everywhere, and two complete coverages are planned.

\subsection{The data}
The CMB is not the only microwave emitter measured by the instruments. Galactic
synchrotron, free-free, and dust emission and extra-galactic sources
contaminate the signal. Fortunately, nearly all of them have emission spectra
which are very different from the CMB. The nine Planck spectral channels cover
one and a half orders of magnitude in frequency. This is necessary to separate
the different physical components observed by a combination of spectral
decomposition and spatial filtering. Detailed properties of the detectors can
be found in Table \ref{tabPlanck}. The resulting physical component maps can
then be analyzed separately according to their nature. Here, only the CMB and
the Galactic emission components are of relevance.

\begin{table*}
\caption{Reproduced from \cite{PlanckBluebook2005}.}
\label{tabPlanck}
\begin{tabular}{lcccccccccc}
\multicolumn{11}{c}{SUMMARY OF PLANCK INSTRUMENT CHARACTERISTICS}\\
\hline
\hline
&\multicolumn{3}{c}{LFI}&&\multicolumn{6}{c}{HFI}\\
\cline{2-4}
\cline{6-11}
INSTRUMENT CHARACTERISTICS&\\    
\hline
Detector Technology&\multicolumn{3}{c}{HEMT arrays}& &
\multicolumn{6}{c}{Bolometer arrays}\\ 
Center Frequency [GHz]&30&44&70&&100&143&217&353&545&857\\
Bandwidth ($\delta \nu/\nu$) &0.2&0.2&0.2&&0.33&0.33&0.33&0.33&0.33&0.33\\
Angular Resolution (arcmin)&33&24&14&&10&7.1&5.0&5.0&5.0&5.0\\
$\delta$T/T per pixel (Stokes I)$^{\rm a}$ &2.0&2.7&4.7& &2.5&2.2&4.8&14.7&147&6700\\ 
$\delta$T/T per pixel (Stokes Q \&U)$^{\rm a}$ &2.8&3.9&6.7& &4.0&4.2&9.8&29.8&$\ldots$&$\ldots$\\
\hline
\end{tabular}\\
$^{\rm a}$ Goal (in $\mu$K/K) for 14 months integration, 1 $\sigma$, for square
  pixels whose sides are given in the row Angular Resolution.
\end{table*}

\section{Primordial magnetic fields}

Particle physics scenarios exist which predict weak primordial magnetic fields,
but usually on very small scales (see papers in these proceedings by
M. Gasperini, by D. Dario, by K. Takahashi, and by D. Sokoloff). K. Subramanian
and also T. Kahniashvili (these proceedings) take the pragmatic approach to
assume that some primordial fields exist and to investigate its observable
signatures. The same approach will be used in the following.

  \subsection{Imprint in power spectra}

A uniform primordial magnetic field would break the cosmic isotropy. Since this
can be tested via CMB data, nG limits on a uniform field were derived from
WMAP and other data, which certainly can be tightened by Planck (see
\cite{2005astro.ph..8544G} for a discussion).

A tangled primordial magnetic field would lead to the presence of Alfv\'en
waves in the primordial plasma --- provided the field is dynamically
significant and there is a non-empty scale interval between the scale of the
field and the ion Larmor radius.  The Alfv\'enic oscillations should imprint
characteristic signatures onto the CMB. On large scales, the development of
these oscillations is suppressed due to the very limited distance the
relatively slow Alfv\'en waves can have traveled for sub-equipartition fields
($B\ll 3 \,\mu$G from constraints of Big Bang nucleosynthesis).  On small
scales, ``oscillations of ... Alfv\'en waves get overdamped in the radiation
diffusion regime, resulting in frozen-in magnetic field perturbations''
(\cite{1998PhRvD..57.3264J}).  This freezing allows the magnetic fluctuations
to survive until recombination and then to produce temperature fluctuations.

{\hskip1mm\hskip-1mm}\cite{1998PhRvL..81.3575S} estimate that fields with
power-law spectra close to scale-invariance of 3~nG result in $10\,\mu$K temperature
fluctuations on multipole scales of $l \sim 1000 - 3000$ (below 10 arcmin). On
sufficiently small scales, the magnetically induced power spectrum will easily
be above the CMB spectrum of the acoustic oscillations due to Silk damping of
the latter. \cite{1998PhRvL..81.3575S} find that for the above parameters this
happens at $l\gtrsim 2000$.

However, the identification of primordial fields using temperature fluctuations
observed by Planck will be challenging. Planck will have an angular resolution
down to 5 arcmin ($l\sim2000$). Therefore, magnetic fields have to be stronger
than assumed above in order for them to exceed the acoustic fluctuations at
these scales and be detectable.  And even if Planck detects excess power at
those scales, a large number of competing explanations is expected.

A unique identification of CMB fluctuations due to magnetic fields may be
possible via the induced polarization signature and/or non-Gaussian statistics.
However, since these are more subtle measurements, their detection in Planck
data is probably also very challenging. We (the authors) do not have a clear
picture of the feasibility of uniquely identifying magnetic contributions to
the CMB fluctuations.

Nevertheless, Planck will be able to constrain primordial magnetic field
scenarios since its measurements can always be used to set upper limits on
magnetically induced fluctuations.

  \subsection{Faraday rotation}

A nice signature of primordial magnetic fields would be an induced Faraday
rotation of the CMB polarization. This effect rotates polarization vectors
with a characteristic $\lambda^2$ dependence, which means that the lowest
frequency channel of Planck will be best suited for such a signal (30 GHz). The
technical specification of Planck requires an accuracy of polarization angle
measurements of about 1 degree. This translates into a sensitivity in rotation
measure of RM$\sim 100\,{\rm rad/m^2}$. 

The Faraday-rotation signal from nG magnetic fields is expected to be about one
degree rotation at 30 GHz (see e.g., \cite{2005PhRvD..71d3006K}). However,
since the CMB polarization is a weak, small-scale signal, the changes due to
Faraday rotation are even weaker and are also on small scales. The expected signal
peaks with $0.1\,\mu$K polarization fluctuations at $l\sim 10^4$. These will be
unobservable for Planck due to beam smearing effects (the 30 GHz system of
Planck will be limited to 33 arcmin resolution $l \sim 400$), and due to the
the overwhelmingly bright and highly polarized Galactic synchrotron emission at
those frequencies extending to high latitudes. Usage of higher frequencies can
decrease the Galactic synchrotron contamination at the price of an even
stronger reduction of the wavelength-dependent Faraday-rotation signal.

In summary, a discovery of primordial magnetic fields via their Faraday
effect is not to be expected from Planck.

\section{Galactic magnetic fields}

The study of primordial magnetic fields with Planck will be difficult,
partly because of the Galactic synchrotron contamination.  However, the
strength of the Galactic foreground signal may open the possibility to study
Galactic magnetic fields in great detail.

  \subsection{Synchrotron, Faraday, and dust polarimetry}

Planck will be sensitive to magnetic fields $\vec{B}$ via
\begin{enumerate}
\item synchrotron emission ($\vec{B}_\perp$),
\item Faraday rotation ($B_\|$),
\item and polarized dust emission ($\vec{B}_\perp/B_\perp$),
\end{enumerate}
where $\perp$ and $\parallel$ refer to the field component with respect 
to the line of sight.

{\bf Synchrotron emission} is ideal for the study of galactic magnetic fields
because the spatial distribution of relativistic electrons illuminating the
fields should be smooth, due to their diffusivity, and can be measured by other
means, e.g., via the inverse Compton scattering of CMB and starlight.

{\bf Faraday rotation} is also ideal because the required spatial distribution of
thermal electrons can be measured by other means (e.g., their free-free
emission). Despite the expected high precision in determining polarization
angles ($\sim 1^\circ$), Planck will only be sensitive to RM $>100\,{\rm
rad/m^2}$ due to its relatively high-frequency channels ($\ge$ 30
GHz). Nevertheless, Planck can provide ``zero-wavelength'' data for a
multi-instrument Faraday campaign.

{\bf Polarized dust emission} due to scattered or partially absorbed starlight
(e.g., \cite{fosalba2001}) or spinning elongated dust grains (e.g.,
\cite{1998ApJ...494L..19D}) might help to build and constrain models of the
galactic field topology. However, it will be difficult to use this for
quantitative investigations of Galactic magnetic fields, due to the complex
dependence of the signal on the poorly known dust properties and spatial
distribution.  

  \subsection{Reconstruction of large-scale fields}

Given the large amount of information that Planck and other polarimetry
instruments will provide on tracers of Galactic magnetic fields, one might ask
if a full reconstruction of the large-scale field based purely on observational
data is possible. The polarimetry of synchrotron emission provides three
observables: the Stokes parameters $I(\lambda)$, $Q(\lambda)$, and
$U(\lambda)$, as functions of wavelength $\lambda$. The goal would be to
reconstruct from these the three magnetic-field components $B_x(z)$, $B_y(z)$,
and $B_z(z)$, which are functions of the coordinate along the line-of-sight
$z$:\\
%
%
\resizebox{\hsize}{!}
{\includegraphics[]{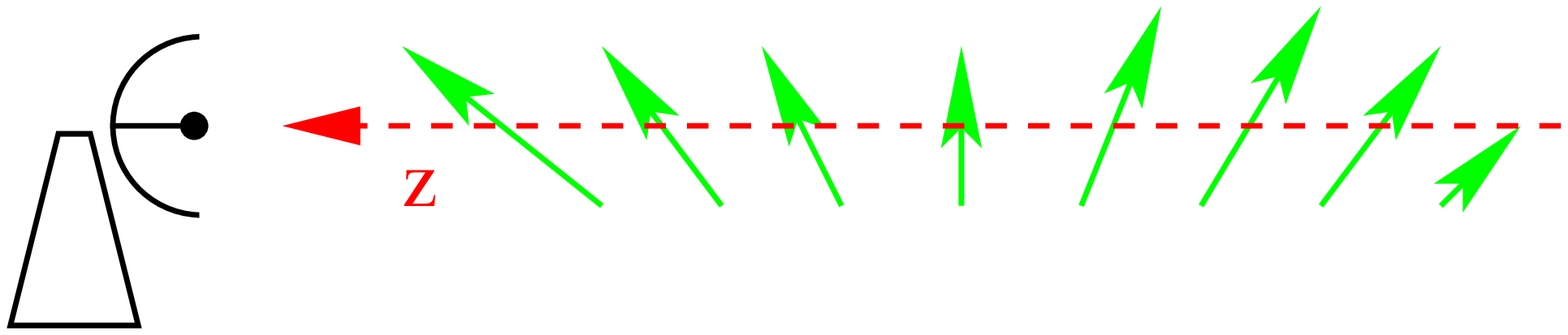}}
{Do we have enough information to reconstruct $\vec{B}(z)$?}

The observed Stokes parameters are given in terms of their emissivities
\begin{eqnarray}
I(\lambda) &=& \int  \! dz\; \eps_I(z,\lambda), \nonumber\\  
P(\lambda) &=& \int  \! dz\; \eps_P(z,\lambda) \, \exp(2 i \lambda^2\,
\phi(z)),\nonumber
\end{eqnarray}
where $P = Q + i U$ denotes the complex polarization, on which the rotation
measure $\phi(z) \propto \int \! dz\,B_z(z)\, n_e(z)$ acts like a mathematical
rotation operator. The emissivities are roughly given by combinations of the
perpendicular magnetic field components:
\begin{eqnarray}
\eps_I &\propto & (B_x^2+B_y^2) \, n_{\rm CRe}, \nonumber\\ 
\eps_P &\propto & (B_x^2-B_y^2 + 2 i B_x B_y) \, n_{\rm CRe}.\nonumber
\end{eqnarray}
The projection of the total intensity emission in the observation removes any
spatial information. For the polarized emission, the frequency-dependent
Faraday effect permits that spatial information imprints itself into the
data. However, introducing the emissivity per Faraday depth
\begin{displaymath}
\eps_{P}(\phi) =
\int \! dz \, \eps_{P}(z) \, \delta(\phi(z) -\phi)  
\end{displaymath}
permits the observable polarization to be expressed without any
reference to the spatial distribution: 
\begin{displaymath}
P(\lambda) = \int^{\infty}_{-\infty}
\!\!\!\!\!\! d\phi \, \eps_P(\phi,\lambda) \, \exp(2 i \lambda^2\, \phi).
\end{displaymath}
 Therefore, a spatial reconstruction is impossible since spatial information is
lost in the projection!\\
%
%
\resizebox{\hsize}{!}
{\includegraphics[]{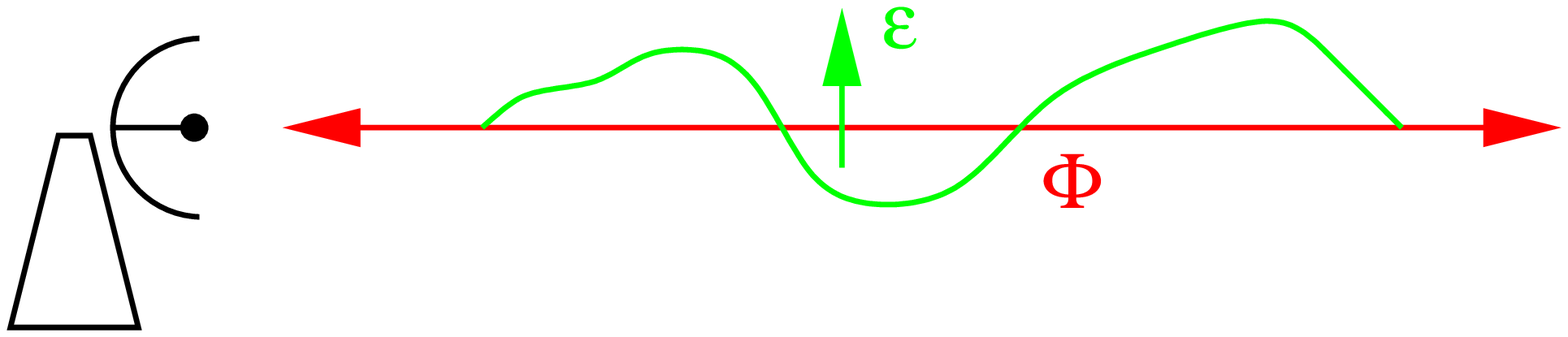}}
{Do we have enough information to reconstruct $\eps_{P}(\phi)$?}

$P(\lambda)$, expressed as a function of the variable $a = 2 \,\lambda^2$, and
$\eps_{P}(\phi)$ are Fourier transforms of each other. Knowing one function
allows the reconstruction of the other. Unfortunately, measurements at negative
values of $a= 2\,\lambda^2$ would be required to reconstruct the complex
function $\eps_{P}(\phi)$, but this is impossible.  Thus, at best half of the
information necessary for a reconstruction of $\eps_{P}(\phi)$ is available.

Although the above equations cannot be inverted, the forward approach is
possible: the construction of polarized emission maps from Galactic models.

\begin{figure}
\resizebox{\hsize}{!}  {\includegraphics[]{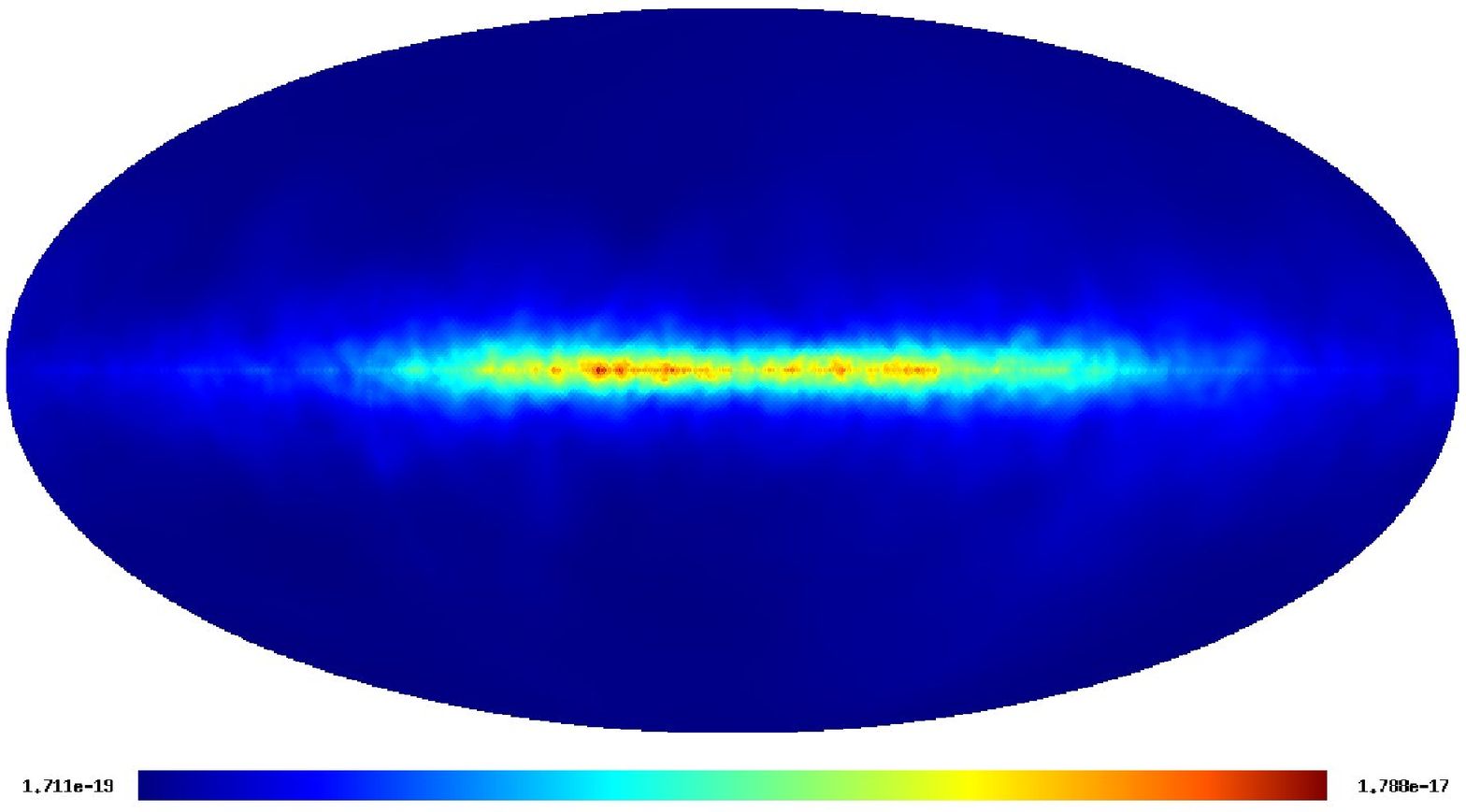}}
\resizebox{\hsize}{!}  {\includegraphics[]{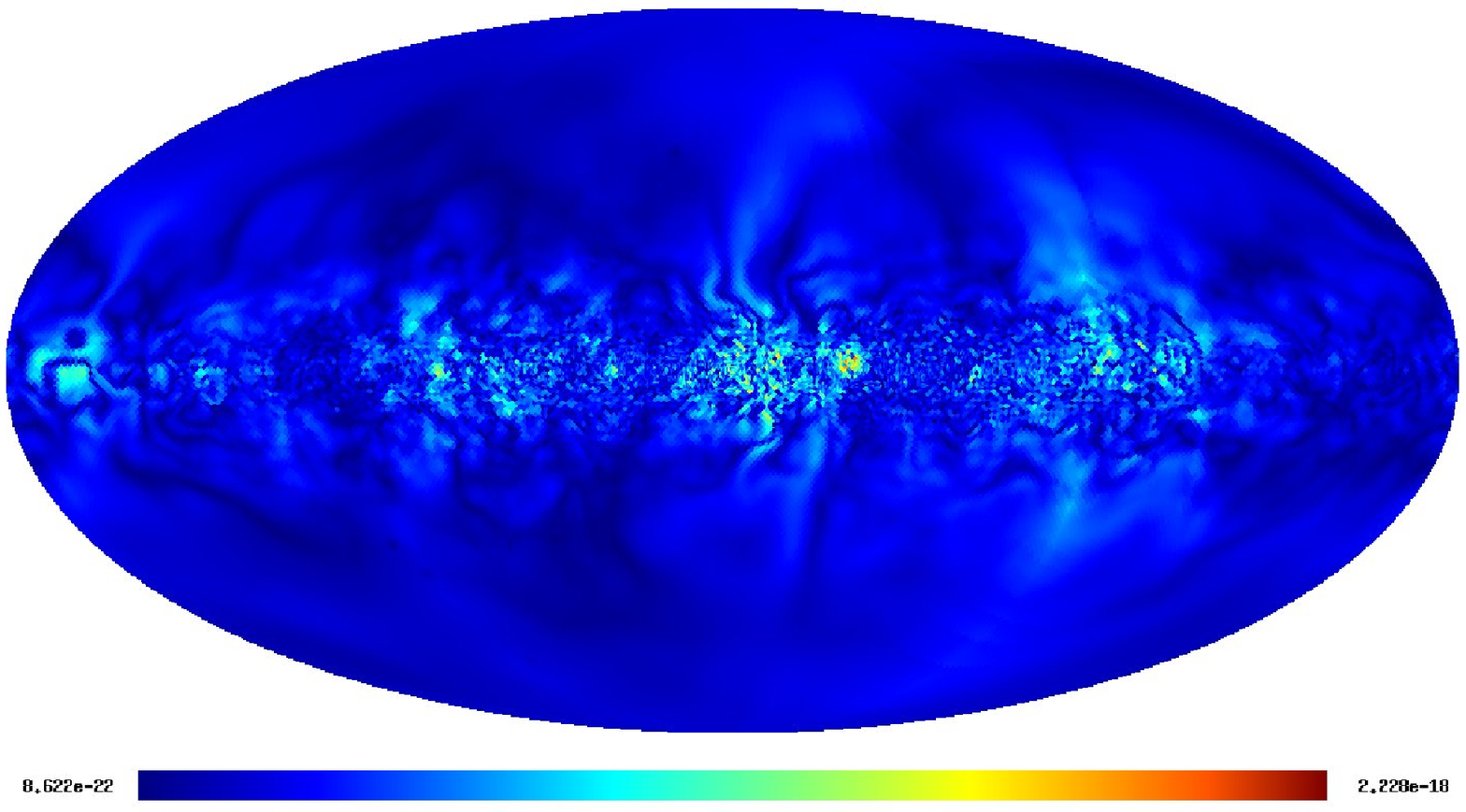}}
\resizebox{\hsize}{!}  {\includegraphics[]{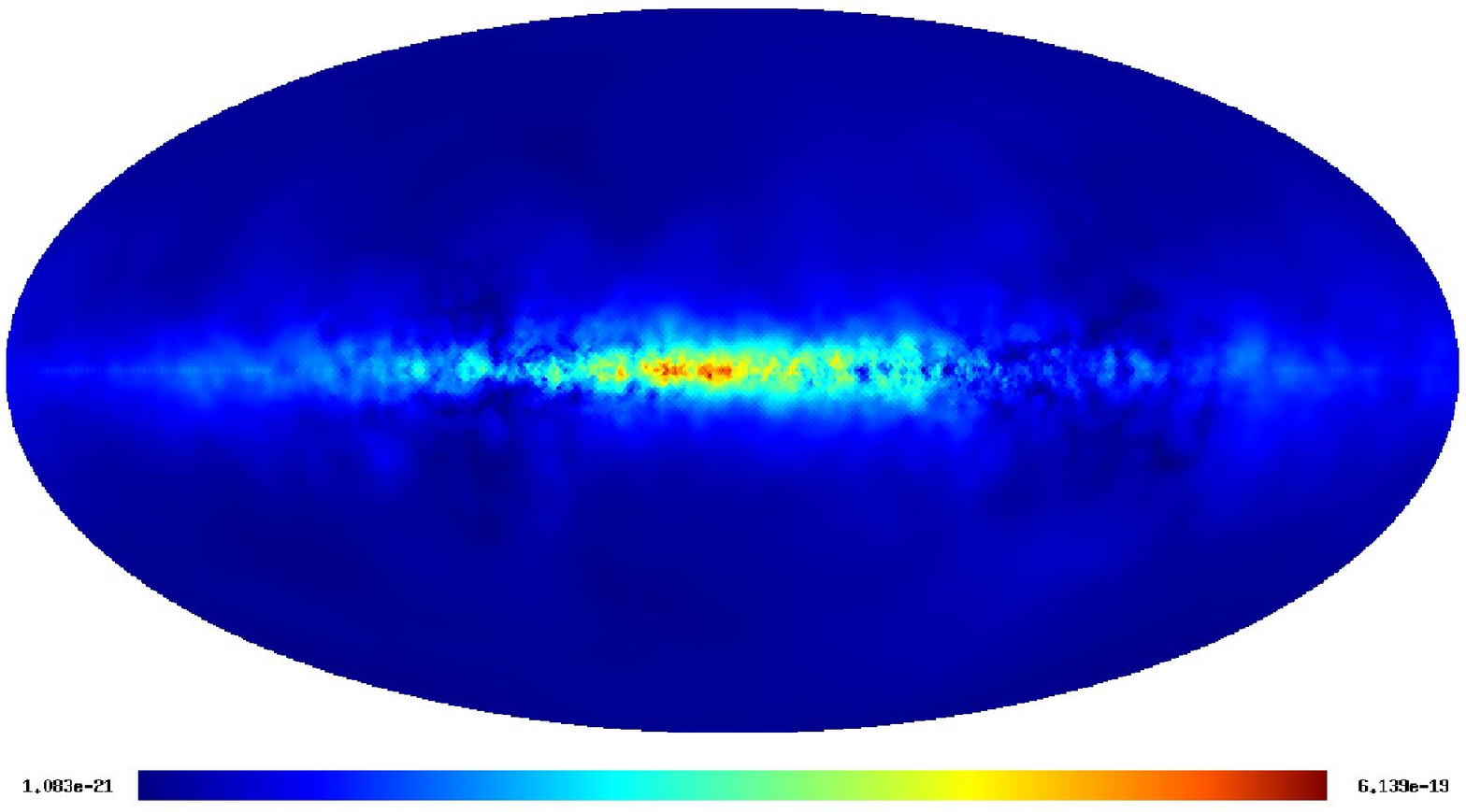}}
\caption{Simulated Galactic total (top) and polarized (middle) intensity at 1.4
  GHz and polarized intensity at the lowest Planck frequency of 30 GHz (bottom)
  calculated with the code of \cite{waelkens05}. Note the Faraday depolarization
  effects in the low frequency map that disappear at high frequencies. 
  The units
  are ${\rm erg/(s\, cm^2\, Hz\, str)}$.}
 \label{fig:Igal}
\end{figure}

The Fourier based formalism of Faraday rotation allows a computationally
efficient calculation of spectral cubes of polarized maps using the fast
Fourier transform. \cite{waelkens05} implemented the generation of maps of
Stokes parameters $I$, $Q$, and $U$ using a nested spherical pixelization
(HEALPix, \cite{2005ApJ...622..759G}) in order to treat beam-depolarization
effects accurately. This code requires as inputs
\begin{itemize}
  \item a Galactic electron model (\cite{2002astro.ph..7156C}),
  \item a Galactic cosmic-ray electron model, 
  \item a Galactic magnetic-field model.
\end{itemize}
Fig. \ref{fig:Igal} shows synthetic emission maps produced by the code for a
very simplistic magnetic field configuration (a logarithmic spiral without
field reversals but with an added random component). Surprisingly, the
simulated maps already reproduce many features of the observed ones, despite
the simplicity of the model used. For instance, the observed Faraday-depolarization
channels, which contribute to the low fractional polarization in the Galactic
plane, seem also to be present in the simulation.

\section{Studying magnetic turbulence}

The Planck-polarization maps may allow a study of properties of MHD turbulence
in the Galaxy with a spatial resolution not achievable in numerical
simulations and a richness in detail not reachable in analytical
investigations.

But how to extract physically meaningful signals? In order to better understand the
technical details of signal extraction, we explain first, how magnetic
power spectra could be extracted from Faraday rotation maps. Then we show that
higher-order statistical properties of the magnetic turbulence may be
measurable from Planck synchrotron-polarization data.

  \subsection{Faraday rotation \& magnetic power spectra}
\begin{figure}
\resizebox{\hsize}{!}
{\includegraphics[width=0.5\textwidth]{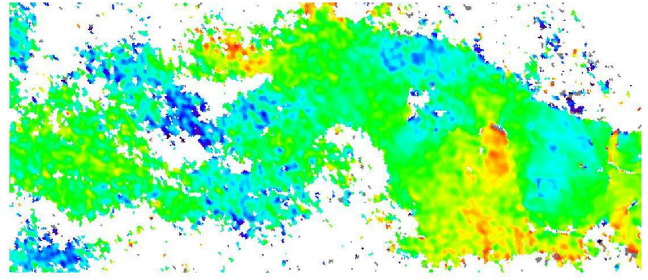}}
{\includegraphics[width=0.5\textwidth]{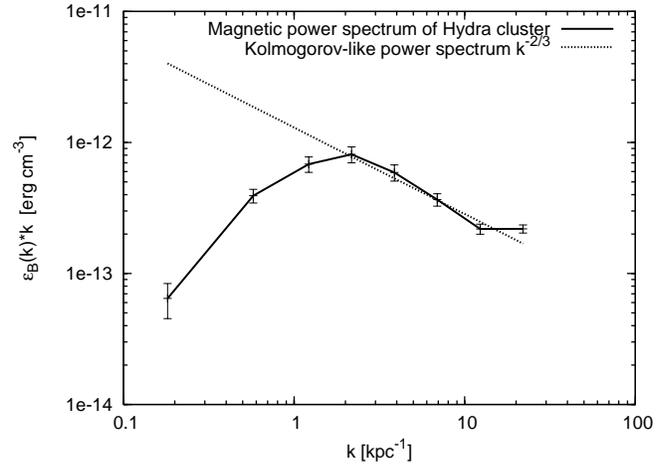}}
\caption{Top: Faraday rotation map of the northern lobe of Hydra A (data:
\cite{1990ApJ...360...41T}, map: \cite{2005MNRAS.358..732V}). 
Bottom: Faraday-based magnetic power spectrum of Hydra A cool core 
(\cite{2005AA...434...67V}). }
 \label{fig:RM}
\end{figure}

Extended and polarized radio sources in galaxy clusters can be used to study
magnetic turbulence in the intergalactic medium. The Faraday rotation map of
Hydra A was analyzed by \cite{2005AA...434...67V} to determine the magnetic power
spectrum within the Faraday screen in front of it (see Fig. \ref{fig:RM}). This
analysis was based on
\begin{itemize}
  \item homogeneous, isotropic, and divergence-free magnetic turbulence,
  \item known cluster geometry (window function), and
  \item Bayes' theorem connecting probabilistically the model spectra 
and the observed RM fluctuations.
\end{itemize}
The incorporation of these assumptions in the analysis is explained in
the following.

Homogeneous magnetic turbulence is best studied using the magnetic-field 
correlation tensor
\begin{displaymath}
\label{eq:defM}
M_{ij}(\vr) = \langle {B}_i(\vx)\,{B}_j(\vx + \vr)\rangle,
\end{displaymath}
whose Fourier space representation is
\begin{displaymath}
\label{eq:defMinFS}
\hat{M}_{ij}(\vk) = \frac{1}{V} \langle \hat{B}_i(\vk)\,\overline{\hat{B}_j(\vk)}
\rangle\,.
\end{displaymath}
In general, this tensor would be described by nine functions defined in the
three-dimensional $k$-space. However, assuming {isotropy} and
$\vec{\nabla}\cdot\vec{B} = 0$, it reduces to
\begin{displaymath}
\label{eq:Mfs}
\hat{M}_{ij}(\vk) =  \frac{1}{2}\, \hat{w}(k) \,\left(\delta_{ij} -
\frac{k_i\,k_j}{k^2}\right) - i \eps_{ijm}\, \frac{k_m}{k}\, \hat{H}(k)\,,
\end{displaymath}
which depends only on two functions of $k=|\vec{k}|$:\\ 
{\it 3-d power spectrum:} 
$\hat{w}(k) = \frac{1}{V} \langle \vec{\hat{B}}(\vk) \vec{\cdot}
\overline{\vec{\hat{B}}(\vk)} \rangle$\\
{\it 3-d magnetic helicity:} $\hat{H}(k) = \frac{i}{2\,V\,k} \langle
\vec{\hat{B}}(\vk) \vec{\cdot} (\overline{\vec{\hat{B}}(\vk)} \times
\vec{k}) \rangle$.

Faraday rotation measures the line-of-sight projected magnetic field
\begin{displaymath}
\RM(\vxp) = a_0 \,\int_{z_{\rm s}(\vxp)}^\infty \!\!\!\!\!\!\! \!\!\!\! dz\;
n_e(\vx) \,B_z(\vx)\,,
\end{displaymath}
where $a_0=e^3/2\pi m_{\rm e}^2 c^4$. 
The RM autocorrelation function
\begin{displaymath}
C_\RM(\vrp) = \langle \RM(\vxp) \,\RM(\vxp+\vrp) \rangle 
\end{displaymath}
is, therefore, 
connected to the magnetic-field correlation tensor. Although we
would need two measurable correlation functions in order to reconstruct the
full tensor, the symmetric part of the tensor is fully encoded in the data
(En{\ss}lin \& Vogt 2003\nocite{2003A&A...401..835E}):
\begin{displaymath}
\hat{C}_\RM^\obs (\vkp) = \frac{1}{2}\,{a_0^2\,n_{\rm
    e,0}^2\,L}\, \hat{w}(|\vkp|)\,.
\end{displaymath}
Here $n_{\rm e,0}$ is a characteristic electron density within the Faraday
screen of depth $L$.  Since the information contained in the power spectrum
alone (and not the helicity) determines the signal, the spectrum can be
inferred from a single observable.

For real data, the situation is a bit more complicated, since the limited size
of the tested volume imprints on the measured correlation function. This can be
described by a window function $W(\vx)$, which contains the geometrical
information on the variation of the cluster electron density and the limited
lateral size of the radio source image. The connection between magnetic-field
and RM spectra becomes:
\begin{displaymath}
\label{eq:CperpwithWindow}
\hat{C}_\RM^\obs (\vkp) = \frac{1}{2}\,{a_0^2\,n_{\rm
    e,0}^2\,L}\, \int \!\!\! d^3q\,\, 
\hat{w}(q)\,\, \frac{q_\perp^2}{q^2}\,
W(\vkp-\vec{q})\,.
\end{displaymath}
This integral equation translating $\hat{w}(q)$ into $\hat{C}_\RM^\obs (\vkp)$
has to be inverted in order to extract the former from the latter.  Bayes'
theorem states that the probability for a model (here the power-spectrum
$\hat{w}(q) $) can be derived from the probability for the data (here the
$\RM$-map) given the model:
\begin{displaymath}
P({\rm model} | {\rm data}) \propto   
P({\rm data}| {\rm model} )\, P({\rm model}). 
\end{displaymath}
Maximizing this probability with respect to the model parameters provides the
Maximum Likelihood power spectrum estimator. This technique
\begin{itemize}
  \item is successfully used in CMB science and was tested for the Faraday
  application with mock RM data,
  \item requires Gaussianity of RM fluctuations, but this is expected
  due to the central limit theorem and is indeed indicated by
  observations,
  \item takes care of the influence of the window function, and
  \item provides errors and an error covariance matrix.
\end{itemize}
It also provides a general conceptual blueprint for extracting statistics of 
physically meaningful quantities from the statistics of the observed signal. 

\subsection{Stokes correlators and Lorentz forces}

Planck will provide high-accuracy and Faraday-free polarization maps.  Are
physically meaningful quantities encoded in them?

\begin{figure}
{\begin{center}
\includegraphics[width=0.5\textwidth]{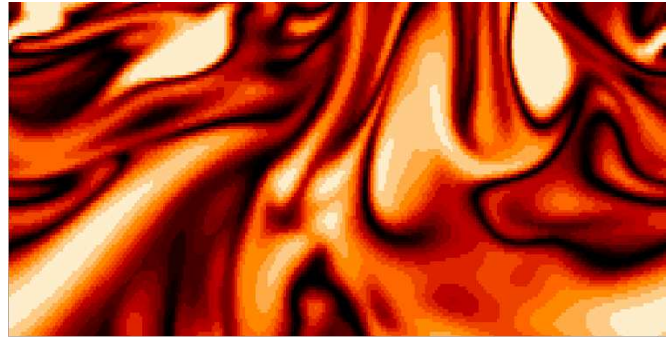}
\end{center}}
\caption{Cross section of the field strength in 
the saturated state of a simulation of homogeneous isotropic 
MHD turbulence (run B in \cite{SCTMM04}).}
 \label{figSchekochihin}
\end{figure}

{\hskip1mm\hskip-1mm}\cite{1989AJ.....98..256E} proposed to use correlation
functions of Stokes parameters (henceforth Stokes correlators) to measure
magnetic power spectra. Since the power spectra are two-point {\em
second-order} statistics, while the Stokes correlators are two-point {\em
fourth-order} statistics, the latter cannot be translated into power spectra
without some closure assumptions.  \cite{1989AJ.....98..256E} demonstrated how
this works for a Gaussian closure.  However, MHD turbulence is not likely to be
well represented by Gaussian fluctuations. This is evident from the appearance
of coherent structures like flux sheets and filaments in numerical simulations
(Fig. \ref{figSchekochihin}) and from the observation of filamentary magnetic
fields in galaxy clusters (Clarke \& En{\ss}lin, these proceedings).

If we were to look for physically meaningful quantities that are directly contained in the
polarized emission data, the most obvious candidate is the Lorentz force 
\begin{displaymath}
\frac{1}{c}\,\vec{J}\times\vec{B}=-{\bf \nabla}\frac{B^2}{8\pi}
+\frac{1}{4\pi}\,\vec{B}\cdot{\bf \nabla}\vec{B}.
\end{displaymath}
Since the Lorentz force is a quadratic quantity, its correlation function 
is a two-point fourth-order statistic. In the above expression, the first term 
on the right-hand side is the magnetic pressure force, the second term is the 
magnetic tension force 
\begin{displaymath}
\vec{F} = \frac{1}{4\pi}\,\vec{B}\cdot\nabla\vec{B},
\end{displaymath}
which not only determines the dynamical response of curved magnetic fields on the 
plasma, but can also be used to diagnose the structure of the field: the 
tension-force correlator measures the gradients of 
the field along itself. Thus, $\langle \vec{F}(\vx) \cdot \vec{F}(\vx + \vr)
\rangle$ can be used as one of the quantitative measures of the 
folded structure evident in Fig. \ref{figSchekochihin}, 
where the field varies across itself on a much larger scale than 
along itself. Further discussion, as well as numerical measurements of the 
tension-force statistics can be found in \cite{SCTMM04} 
(see also the review by \cite{SC05}). 
We propose to study the statistics 
of the tension force using polarized radio observations. 

The correlation tensor of the tension-force 
or its Fourier counterpart
\begin{eqnarray*}
\frac{1}{V}\bigl\langle \hat{F}_i(\vec{k})\overline{\hat{F}_m(\vec{k})}\bigr\rangle
&=& k_j k_n \hat{C}_{ij,mn}(\vec{k})
\end{eqnarray*}
where $\hat{C}_{ij,mn}(\vec{k})$ is the Fourier-space representation of  
the two-point fourth-order magnetic correlation tensor
\begin{displaymath}
 C_{ij,mn}(\vec{r})=\langle B_i(\vec{x}) B_j(\vec{x})
 B_m(\vec{x}+\vec{r}) B_n(\vec{x}+\vec{r})\rangle.
\end{displaymath}
A general isotropic fourth-rank tensor 
depends on 26 scalar functions $C_{1\ldots26}(r)$ of the distance $r=|\vec{r}|$. 
Fortunately, tensor symmetries allow 
this set to be reduced to seven unknown scalar functions that fully determine 
the two-point fourth-order statistics of the field. In Fourier 
space, this reads
\begin{eqnarray*}
&&\!\!\!\!\!\hat{C}_{ij,mn}(k) = \hat{C}_1(k)\delta_{ij}\delta_{mn} +
\hat{C}_2(k)(\delta_{im}\delta_{jn}+\delta_{in}\delta_{jm})\\ 
\ &&+\ \hat{C}_3(k)\hat{k}_i \hat{k}_j \hat{k}_m \hat{k}_n 
+ \hat{C}_4(k)(\delta_{ij}\hat{k}_m\hat{k}_n+\delta_{mn}\hat{k}_i \hat{k}_j)\\
\ &&+\ \hat{C}_5(k)(\delta_{im}\hat{k}_j \hat{k}_n+\delta_{in}\hat{k}_j\hat{k}_m
+ \delta_{jm}\hat{k}_i \hat{k}_n+\delta_{jn}\hat{k}_i\hat{k}_m)\\ 
\ &&+\ i\hat{C}_6(k)(\epsilon_{imp}\hat{k}_p\delta_{jn}+\epsilon_{inp}\hat{k}_p\delta_{jm}\\
\ &&\quad\quad\quad\ +\,\epsilon_{jmp}\hat{k}_p\delta_{in}+\epsilon_{jnp}\hat{k}_p\delta_{im})\\
\ &&+\ i\hat{C}_7(k)(\epsilon_{imp}\hat{k}_p\hat{k}_j\hat{k}_n+\epsilon_{inp}\hat{k}_p\hat{k}_j\hat{k}_m\\
\ &&\quad\quad\quad\ +\,\epsilon_{jmp}\hat{k}_p\hat{k}_i\hat{k}_n+\epsilon_{jnp}\hat{k}_p\hat{k}_i\hat{k}_m), 
\end{eqnarray*}
where $\hat{C}_{1\ldots7}(k)$ are all real. 
Is it possible to reconstruct these seven functions from observations? The
accessible observables are the Stokes parameters.  These are related to the
magnetic fields in the emission region for Faraday-free, steep spectrum
($\alpha \approx -1$) radio emission:\footnote{Here constant factors converting
field strength to radio emissivity have been suppressed for simplicity of the
calculations.}
\begin{eqnarray*}
I(\vec{x}_\perp) & = & \int_{z_0}^\infty \!\!\! dz \left[ B_x^2(\vec{x})+ 
B_y^2(\vec{x})\right]\\
Q(\vec{x}_\perp) & = & \int_{z_0}^\infty \!\!\! dz \left[ B_x^2(\vec{x})- 
B_y^2(\vec{x})\right]\\
U(\vec{x}_\perp) & = & \int_{z_0}^\infty \!\!\! dz \,2 \, B_x(\vec{x}) 
B_y(\vec{x}),
\end{eqnarray*}
from which we can construct six scalar Stokes correlators: in Fourier space, 
$\hat{\Sigma}_{II}(\vec{k}_\perp)$, $\hat{\Sigma}_{QQ}(\vec{\vk}_\perp)$,
$\hat{\Sigma}_{UU}(\vec{k}_\perp)$, $\hat{\Sigma}_{IQ}(\vec{\vk}_\perp)$,
$\hat{\Sigma}_{IU}(\vec{k}_\perp)$, and $\hat{\Sigma}_{QU}(\vec{k}_\perp)$. 
These can be expressed as linear combinations of the scalar functions 
$\hat{C}_{1\ldots5}$, while $\hat{C}_6$ and $\hat{C}_7$ remain indeterminable. 
It turns out that the Stokes correlators are not 
independent and only four scalar correlation functions 
are, in fact, available from observations. 
Thus, we are short by one such function to fully reconstruct $\hat{C}_{1\ldots5}$. 

Despite this scarcity of observable information, 
the power spectrum of the tension force happens to be fully observable!
It is completely expressed in terms of the Stokes correlators:
\begin{eqnarray*}
{1\over V}\bigl\langle \vec{\hat{F}}(\vec{k}_\perp) \vec{\cdot}
\overline{\vec{\hat{F}}(\vec{k}_\perp)} \bigr\rangle
=
\frac{\vec{k}_\perp^2}{8\pi} 
\int_0^{2\pi}\!\!\!d\varphi\left[\hat{\Sigma}_{II}(\vec{k}_\perp) 
+ 3\hat{\Sigma}_{QQ}(\vec{k}_\perp)\right.\\
\left.-\left(\hat{\Sigma}_{QQ}(\vec{k}_\perp) 
- \hat{\Sigma}_{UU}(\vec{k}_\perp)\right)\cos 4\varphi 
+ 4\hat{\Sigma}_{IQ}(\vec{k}_\perp)\cos2\varphi\right],
\end{eqnarray*}
where $\varphi$ is the angle between $\vec{k}_\perp$ and the $x$ axis.  This
demonstrates that physically relevant information on MHD turbulence is encoded
in the Stokes correlators and, therefore, can be obtained from Planck
polarization data! The feasibility of this approach and the reliability of the
underlying assumptions clearly require further careful investigation, which is
currently underway.

\section{Conclusions}

\begin{enumerate}
\item {The Planck surveyor mission} is a high-precision experiment to study
  cosmology with the CMB.
\item However, an unambiguous detection with Planck of primordial magnetic
  fields --- both from their imprint in the CMB power spectra and from their
  Faraday rotation --- will be extremely challenging.  But interesting
  constraints should be possible.
\item Galactic magnetic fields can well be studied by Planck via 
synchrotron, Faraday, and dust polarimetry, allowing a model-based reconstruction of 
large-scale fields.
\item {Studying magnetic turbulence} with Planck is promising: a new technique
  of Stokes correlators may allow us to measure the Lorentz-force power spectra
  in MHD turbulence, similar to the Faraday-based magnetic power spectra
  estimates.
\end{enumerate}

\acknowledgements TAE thanks the SOC for the invitation to and the LOC for the
warm hospitality at this exciting conference.  This work has benefited from
research funding from the European Community's sixth Framework Programme under
RadioNet R113CT 2003 5058187. AAS was supported by the UKAFF 
Fellowship.

\end{document}